\documentclass[aps,prb,superscriptaddress,groupedaddress]{revtex4-1}
\usepackage{graphicx}  % needed for figures
\usepackage{dcolumn}   % needed for some tables
\usepackage{bm}        % for math
\usepackage{amssymb}   % for math
\usepackage{amsmath}   % for math
\usepackage{subfigure}
\usepackage{gensymb}
\newcommand{\mr}{\mathrm}

\begin{document}

% The following information is for internal review, please remove them for submission
\widetext
%\leftline{Version 2 as of \today}

\title{Molecular Polarizability of Water from the Local Dielectric Response Theory}

\author{Xiaochuan Ge}
\altaffiliation{Present address: BlackRock, Inc.}
\author{Deyu Lu}
\email{dlu@bnl.gov}
\affiliation{Center for Functional Nanomaterials, Brookhaven National Laboratory
, Upton, New York 11973, United States}
\date{\today}

\begin{abstract}
We propose a fully \emph{ab initio} theory to compute the electron density response under the perturbation in the local field.
This method is based on our recently developed local dielectric response theory~[Phys. Rev. B  92, 241107(R), 2015], which provides
a rigorous theoretical framework to treat local electronic excitations in both finite and extended systems beyond the commonly employed dipole approximation.
We have applied this method to study the electronic part of the molecular polarizability of water in ice I\emph{h} and liquid 
water. Our results reveal that the crystal field of the hydrogen-bond network has strong anisotropic effects, which significantly enhance the out-of-plane component and
suppress the in-plane component perpendicular to the bisector direction. The contribution from the charge transfer is equally important,
which increases the isotropic molecular polarizability by $5-6$\%. Our study provides new insights into the dielectric properties of water,
which form the basis to understand electronic excitations in water and to develop accurate polarizable force fields of water.

\end{abstract}

\maketitle
% Introdcution
\section{Introduction}
Water is one of the most important substances for life,  many fields of science, and numerous technological applications.
Despite its simple molecular structure, water has many anomalous behaviors, e.g., its density reaching the maximum at 4\degree C, that have attracted  
intensive research for decades. One intriguing aspect of the fundamental properties of water is the electronic excitation, which is essential to the understanding of
a broad range of problems, such as solvation, water/solid interfaces, and electrochemical reactions in liquid solution.
The quantum theory of electronic excitation has been widely used to interpret, e.g., the x-ray absorption~\cite{CAVA2002,HETE2004,PREN2006,CHEN2010} and optical absorption
spectra of water~\cite{HAHN2005,GARB2006}, which in turn provide important physical insights into the atomic structures and the electronic structure of water. 
Besides, it has been well established that van der Waals dispersion forces, arising from the coupling of instantaneous and induced excitations, play a critical role in accurately describing the structure of water in \emph{ab initio} molecular dynamics simulations~\cite{SANT2008,ZHAN2011,SANT2011,WANG2011,GILL2016}. 

In the linear response regime, how to describe the electronic part of the molecular polarizability of water ($\alpha_{\rm{H_2O}}$) remains a subject of debate, despite the rather simple textbook picture of dielectrics~\cite{KITT96}. Experimentally, $\alpha_{\rm{H_2O}}$ is known in the gas phase only~\cite{MURP77}. In the condensed phase, the average value of the molecular polarizability is typically estimated from the refraction index measurement using the Lorentz-Lorenz relation~\cite{DEBY12}. Although this approach is suitable for gases and nonpolar liquids, it fails for polar liquids, such as water, due to the inaccurate description of the local field experienced by the polar molecule. Many competing models have been proposed to improve Debye's dipole theory~\cite{DEBY12}, including models from Onsager~\cite{ONSA36} and Kirkwood~\cite{KIRK39}. Due to the deficiency of a classical treatment of the local field in liquid water,
this problem has not been solved yet. Besides the actual value of $\alpha_{\rm{H_2O}}$, it is more important to understand how $\alpha_{\rm{H_2O}}$ changes in different chemical environments (e.g. from the gas phase to the condensed phase) or under different boundary conditions (e.g., bulk water or confined water, such as water at the solid / water interface). 

From the computational point of view, although the macroscopic average of the dielectric response of water is very well captured by the density functional theory (DFT) or beyond-DFT methods, such as the many-body perturbation theory~\cite{HAHN2005,GARB2006}, the chemical nature of the microscopic counterpart, e.g., the microscopic electric susceptibility ($\chi$), remains poorly understood. Currently, a rigorous, fully \emph{ab initio} theory to compute $\alpha_{\rm{H_2O}}$ is lacking, and existing models predict contradicting trends in 
$\alpha_{\rm{H_2O}}$. Using point charge models to represent the solvent molecules, Gubskaya and Kusalik~\cite{GUBS2001} found an increase of $\alpha_{\rm{H_2O}}$ in liquid.
Morita~\cite{MORI2002} proposed a cluster model, where the solute polarizability in the cluster is approximated by the difference of the polarizability of the cluster
and that of the solvent only, $\Delta \alpha=\alpha^{\rm{tot}}-\alpha^{\rm{solv}}$. The non-additive correction was taken into account by introducing a dielectric continuum model of the solvation shell~\cite{MORI2002}. In contrast to the point charge model, the cluster model predicts the isotropic molecular polarizability, $\bar{\alpha}_{\rm{H_2O}}=\frac{1}{3}\rm{Tr}\,(\alpha_{\mr{H_2O}})$,  in the liquid is reduced from that in the gas phase by $7-9$\%, and the reduction was attributed to the electron repulsion of the ambient solvent molecules that perturb and confine the spatially diffuse tail of the electron cloud of the solute~\cite{MORI2002}. 

Most of other computational studies employed an extension~\cite{HEAT2006} of the interactive dipole model (IDM)~\citep{SILB17A, *SILB17B, *SILB17C,APPL72,THOL81} to the condensed phase, where the electrostatics of the electron density response is approximated at the dipole-dipole interaction level. IDM calculations found that $\bar{\alpha}_{\rm{H_2O}}$ in the liquid water is reduced by less than $2$\%~\cite{SALA2008,BUIN2009} or the same as~\cite{WAN2013} that of the gas phase. These results also showed an anisotropic effect, with the in-plane components reduced, while the out-of-plane component enhanced~\cite{SALA2008,BUIN2009,WAN2013}. The origin of these changes is unclear, as the effects of the crystal field are intertwined with structural changes of water monomer geometries under the thermal fluctuation. 

Because of the uncertainties in determining $\alpha_{\rm{H_2O}}$ from both experiment and theory, there is no clear recipe on how to choose the \emph{right} $\alpha_{\rm{H_2O}}$ in a polarizable force field, given the knowledge of the gas phase value~\cite{POND2010}. To address this open question, in this work we proposed a fully \emph{ab initio} method to compute $\alpha_{\rm{H_2O}}$ based on the our recently developed local dielectric response theory~(LDRT)~\cite{GE2015}, which is among the techniques to compute the dielectric response function without explicitly referring to the empty states~\cite{BARO87,Baroni2001,WILS2008,WILS2009,WALK2006,ROCC2008,ROCC2010}. The new method is general and can be used to study $\alpha_{\rm{H_2O}}$ in different chemical environments. We proved that the widely used IDM is the dipole limit of our theory. In the numerical study of $\alpha_{\rm{H_2O}}$ in liquid water, we paid special attention to separate the environmental effects due to the crystal field and charge transfer from those caused by intramolecular thermal fluctuations. 

\section{Method}
\subsection{Molecular polarizability in extended systems under the dipole approximation}
In order to quantify molecular polarizabilities in an extended system,
it is necessary to express the dielectric response of a many-electron
system in a local representation, instead of the Bloch
representation. A formal way to proceed is to use the Wannier function
(WF) formalism, such as the maximally localized Wannier function
(MLWF)~\cite{Marzari1997,SOUZ2001,Marzari2012}, where the WFs ($|w_{{\bf R}n}\rangle$) are constructed
from unitary transformations of the Bloch orbitals ($|\psi_{m}^{\bf
k}\rangle$). In systems with a finite band gap, we consider only the
occupied bands:
\begin{equation}
|w_{{\bf R}n}\rangle =
  \frac{\Omega}{(2\pi)^3}\int_{BZ} d{\bf k} e^{-i{\bf k}\cdot {\bf R}}
  \sum_{m=1}^{n_v} U^{({\bf k})}_{mn} |\psi_{m}^{\bf k} \rangle,
\end{equation}
where $n_v$ is the total number of the occupied bands, and $\Omega$ is
the real space primitive cell volume. Unitary matrices ($U^{({\bf
k})}$) in the MLWF formalism minimize the spatial spreads of the WFs
labeled by the lattice vector (${\bf R}$) and the Wannier center
index ($n$). In this study, we focus on isolated systems or periodic
systems with a large supercell, where a single $\Gamma$-point
sampling in the reciprocal space is used. The extension to general
cases of the $k$-point sampling, although in principle feasible, is
beyond the scope of this work. In the $\Gamma$-point formulation, the
above expression is simplified to
\begin{equation}
|w_n\rangle =\sum_{m=1}^{n_v} U_{mn} |\psi_{m} \rangle.
\end{equation}
Within the modern theory of the electric polarization in crystalline
dielectrics~\cite{KING93,VAND93,SOUZ2000,REST1994}, the dipole moment of a sub-system (e.g., a ion or
molecule labeled by $M$) is defined in atomic units as
\begin{equation}
\bm{\mu}_M=\sum_{i\in M} Z_i\bm{R}_i^{ion} -2 \sum_{n\in M} \bm{r}_n,
\end{equation}
where $Z_i$ and $\bm{R}_i^{ion}$ are the charges and positions of ions.  
Positions of the Wannier centers that belong to $M$,  
$\bm{r}_n=x_n\hat{i}+y_n\hat{i}+z_n\hat{k}$, can be computed from~\cite{Marzari1997}
\begin{equation}
x_n=-\frac{L}{2\pi}\Im\ln\langle w_n|e^{-i\frac{2\pi}{L}x} |w_n \rangle,
\end{equation}
where $L$ is the size of the supercell.

Upon a small perturbation from an external electric field ($\bm{E}$)
at a given frequency, the electronic contribution to the net induced
dipole of $M$ in the gas phase is given by
\begin{equation}
\Delta\bm{\mu}_M=\alpha_M^{gas}\;\bm{E}, \label{Ealaphgas}
\end{equation}
where $\alpha_M^{gas}$ is the frequency-dependent gas phase molecular
polarizability tensor. Here we dropped the explicit frequency
dependence to simplify the notation. Once the molecule is embedded in
a chemical environment, e.g. a cation in a solid or a solute
molecule surrounded by solvent molecules, the induced dipole arises
from the response to the local field that is the sum of the external
field and the induced field from the environment,
$\bm{E}_M^{loc}=\bm{E}+\bm{E}_M^{ind}$. Unlike $\bm{E}$,
$\bm{E}_M^{ind}$ is a microscopic quantity, whose direction and
amplitude vary within the size of the molecule. In practice, this
finite size effect is often ignored, which leads to the approximate
expression,
\begin{equation}
\Delta\bm{\mu}_M=\alpha_M\;\bm{E}_M^{loc}, \label{Edmu}
\end{equation}
where $\alpha_M$ is the environment-dependent molecular
polarizability tensor.

Under the dipole approximation, $\bm{E}_M^{ind}$ in the IDM is approximated by
the dipole field of environment molecules ($N\neq M$),
\begin{equation}
\bm{E}_M^{ind}\approx\sum_{N\neq M} T_{MN}\; \Delta\bm{\mu}_N, \label{EEind}
\end{equation}
where $T_{MN}=\nabla\nabla r_{MN}^{-1}$ is the dipole-dipole
interaction tensor with $r_{MN}$ the inter-molecular
distances. Substituting Eq.~\ref{EEind} into Eq.~\ref{Edmu}, one can
derive an equation of interacting dipoles~\citep{SILB17A, SILB17B, SILB17C,APPL72,THOL81,HEAT2006},
\begin{equation}
\Delta\bm{\mu}_M=\alpha_M \; (\bm{E}+\sum_{N\neq M} T_{MN}\; \label{Edipole} \Delta\bm{\mu}_N).
\end{equation}
$\alpha_M$ in Eq.~\ref{Edipole} can be solved conveniently, once 
$\Delta\bm{\mu}$ are obtained from finite external field calculations~\cite{HEAT2006,BUIN2009}
or linear response theory~\cite{WAN2013}. 

\subsection{Local dielectric response theory}
\begin{figure}[tbh]
\includegraphics[width=2.5 in]{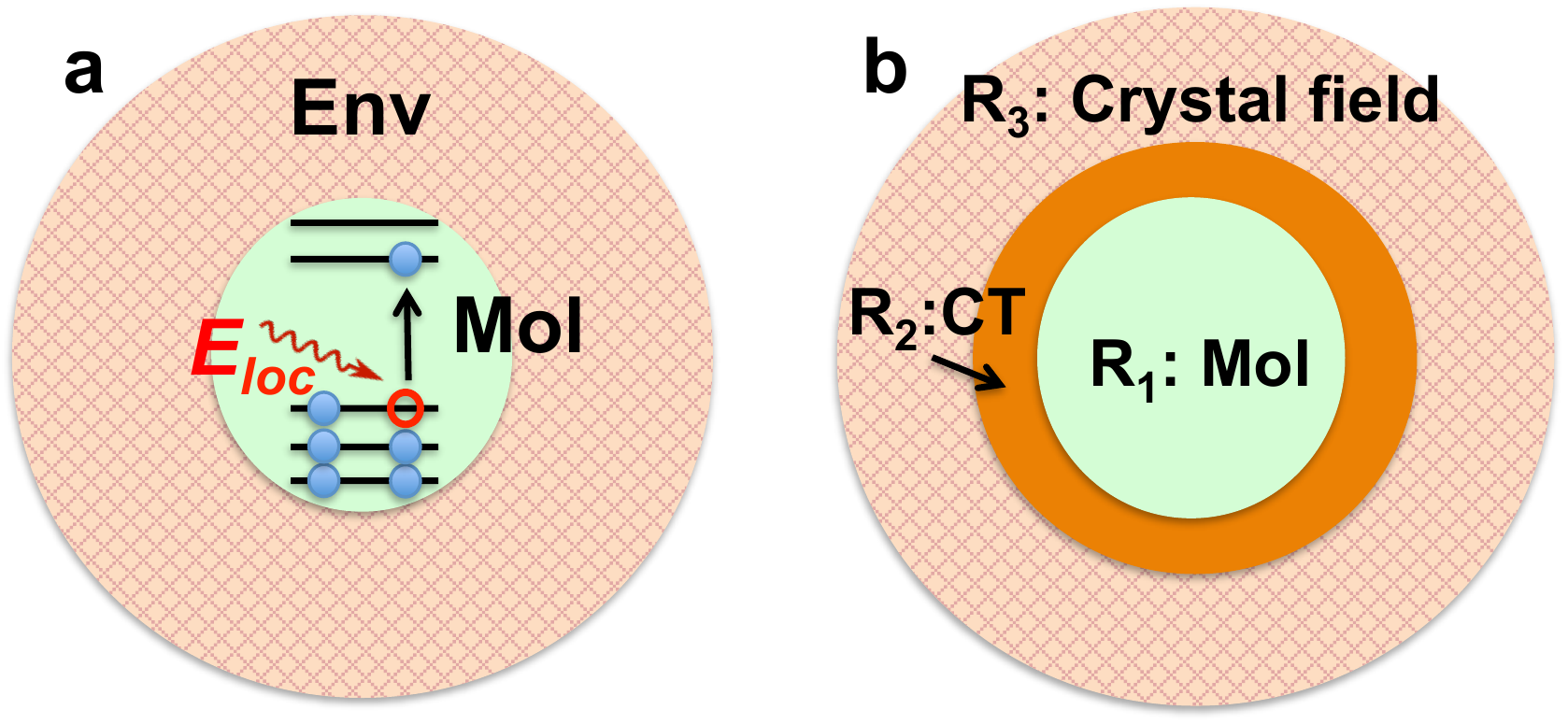}
\caption{ \label{fig-lext} a) Schematics of local excitations of a
molecule inside a chemical environment. b) Three regions in the
quantum mechanical description of the problem, R$_1\in$ R$_2\in$ R$_3$. R$_1$: the region of the
molecule of interest, R$_3$: the whole system providing the
crystal field (CF), and R$_2$: a subset of R$_3$ coupled to R$_1$ through
charge transfer (CT) upon local excitations.}
\end{figure}

Although the IDM combined with DFT has been widely used to calculate
molecular polarizabilities~\cite{HEAT2006,SALA2008,BUIN2009,ROTE2010,WAN2013}, it has several
drawbacks. First of all, as mentioned above, Eq.~\ref{Edmu} neglects the finite size
effect in the induced field of the environment. Secondly, the dipole approximation neglects all the higher order terms. It might be justifiable at the
far field, but when molecules get close to each other, the validity of the dipole 
approximation is questionable. More severely, the use of $T_{MN}$ does not satisfy the 
positive definite requirement of Eq.~\ref{Edipole}~\cite{THOL81}. As a consequence, molecular 
polarizabilities from the IDM can diverge or become negative in simple systems, e.g., the 
cooperative (head to tail) induced dipoles in the direction of the line connecting the 
two~\cite{THOL81, VAND98, ALLE2004}. Finally, at the interface between $M$ and the environment (see Fig.~\ref{fig-lext}b),
wavefunctions of excited environmental electrons can have finite overlap with the occupied valence
electron orbitals of $M$, leading to a charge transfer (CT) type of density response on $M$.
The CT contribution is a quantum effect, and can not be captured at the level of dipole-dipole interaction.

In order to develop a rigorous, fully quantum mechanical treatment of $\alpha_M$, let us first 
consider a slightly broader scenario, the local excitations of a molecule embedded a chemical environment
as shown in Fig.~\ref{fig-lext}a. The whole system can thus be divided into three regions (R$_1\in$ R$_2\in$ R$_3$) as shown in 
Fig.~\ref{fig-lext}b, R$_1$: the region of the molecule of interest ($M$), R$_3$: the whole system providing the
crystal field (CF), and R$_2$: a subset of R$_3$ coupled to R$_1$, such that
CT can happen between R$_2$ and R$_1$ upon local excitations. We will introduce a formal
theory based on the microscopic susceptibility, $\chi$. We note that Allen~\cite{ALLE2004} developed a model of 
charge-dipole interaction along the same line, where charges are treated by quantum mechanics, and charge interactions
are solved in the random phase approximation (RPA). The remaining
obstacle in our approach is to reformulate $\chi$, a non-local quantity by definition,
in the local representation. 

An early proposal was made by Hanke to address this issue by localizing electrons and holes separately when 
building an explicit electron-hole pair basis~\cite{HANK73}. In practice, this method is inconvenient, because
of a) the difficulty to numerically converge the number of unoccupied bands, and b) even if the convergence can be reached, the poor locality of
the high energy unoccupied bands. Both limitations of Hanke's method~\cite{HANK73} can be avoided, because it is unnecessary to localize the
unoccupied bands. In other words, we formulate the theory on the occupied manifold only, without using the explicit electron-hole pair basis.
In the following, we first summarize the recently developed local dielectric response theory (LDRT)~\cite{GE2015} that
provides the theoretical framework to study local excitations. Then we
use the LDRT to develop a fully \emph{ab initio} theory to calculate the
dielectric response of the perturbation in the local field.

The central quantities in the linear response theory are the bare,
$\chi^0(\omega;\bm{r},\bm{r}')$, and the screened susceptibility,
$\chi(\omega;\bm{r},\bm{r}')$, which are the functional derivatives of
the charge density response with respect to perturbations in the
self-consistent field potential ($\delta V_{scf}$) and the external
potential ($\delta V_{ext}$):
\begin{eqnarray}
\chi^0(\omega;\bm{r},\bm{r}')&=\delta \rho(\omega;\bm{r}) /\delta V_{scf}(\omega;\bm{r}'), \nonumber \\
\chi(\omega;\bm{r},\bm{r}')&=\delta \rho(\omega;\bm{r}) /\delta V_{ext}(\omega;\bm{r}').
\end{eqnarray}
In the following, we adopt the shorthand notation: $\Delta \rho =
\chi^0\, \Delta V_{scf}=\chi\, \Delta V_{ext}$, where integration on
common variables is implied. $\chi$ can be solved from $\chi^0$
through Dyson's equation~\cite{FETT2003}, $\chi = \chi^0 + \chi^0 \,
K\, \chi$, where $K=v_H + K_{xc}$ with $v_H$ and $K_{xc}$ being the
Coulomb and exchange-correlation kernel, respectively. In the language of the linear response theory,
$\chi^0$ is the building block.

Under the Bloch representation, both $\Delta\rho$ and $\chi^0$ can be partitioned according to occupied bands, 
\begin{eqnarray}
\Delta\rho &=& \sum_v \Delta \rho_v, \;\;\;\;\;\;\;\;\;\;\;\;\;\;\;\;\;\;\;\;\;\;\; \chi^0 = \sum_v \chi^0_v, \nonumber \\
\Delta \rho_v &=& \chi^0_v\;\Delta V_{scf}=2\, (|\Delta \psi_{v}^+ \rangle+|\Delta \psi_{v}^- \rangle) \langle \psi_v|,
\end{eqnarray}
where $|\Delta \psi_v^\pm \rangle$ are the solution of the Sternheimer
equation at $+$ and $-$ frequencies~\cite{Baroni2001},
\begin{equation}
  (\varepsilon_v-\hat{H}-\alpha \hat{P} \pm \omega) \left | \Delta \psi_v^\pm \right 
  \rangle = \hat{Q} \,\Delta V_{scf} \left | \psi_v \right \rangle.
  \label{stern-KS}
\end{equation}
Here $\varepsilon_v$ are energy
levels of the occupied states of the Kohn-Sham (KS) Hamiltonian ($\hat{H}$);
$\hat{P}=\sum_v \left |\psi_v \right \rangle \langle \psi_v |$ and
$\hat{Q}=\hat{I}-\hat{P}$ are projectors onto the occupied and unoccupied state
manifolds, which are introduced to avoid the explicit reference to the
unoccupied states~\cite{Baroni2001}. The term $\alpha \hat{P}$ is introduced
to remove the singularity of Eq.~\ref{stern-KS}.

The essence of the LDRT is to recognize that $\Delta \rho$ is invariant
under the unitary transformation of the occupied orbitals, which
allows $\Delta \rho$ and $\chi^0$ to be formulated under the Wannier
representation,
\begin{equation}
\Delta \rho=\sum_n 2\, (|\Delta W_{n}^+ \rangle+|\Delta W_{n}^- \rangle) \langle W_n|,
\end{equation}
where $|\Delta W_n^\pm\rangle$ are solutions of the generalized Sternheimer equation~\cite{GE2015},
\begin{equation}
\sum_{n'}(\tilde \varepsilon_{nn'}-\hat{H}-\alpha \hat{P} \pm \omega) \left 
| \Delta W_{n'}^\pm \right \rangle = \hat{Q} \Delta V_{scf} \left | W_{n} \right \rangle .
  \label{stern-WF}
\end{equation}
Since $|W_n \rangle$ are not the eigenstates of the KS Hamiltonian,
$\varepsilon_v$ are replaced by the coupling matrix elements, $\tilde
\varepsilon_{n,n'}=\langle W_n|H|W_{n'}\rangle$. In contrast to
Eq.~\ref{stern-KS}, linear response equations in Eq.~\ref{stern-WF}
are entangled due to $\tilde \varepsilon_{n,n'}$. The variation of
$|W_{n}\rangle$ caused by the perturbation at $|W_{n'}\rangle$ can be
obtained from
\begin{equation}
\begin{aligned}
\left | \Delta W_{n}^\pm \right \rangle &= \sum_{n'} \left | \Delta W_{nn'}^\pm \right \rangle \\ 
&\equiv \sum_{n'} [\tilde \varepsilon -(\hat{H}+\alpha \hat{P}\mp \omega)\,\hat{I}]^{-1}_{nn'} \,\hat{Q} \,\Delta V_{scf} \left | W_{n'}\right \rangle , \label{stern-WF-sol}
\end{aligned}  
\end{equation}
where $\hat{I}$ is an $N_v \times N_v$ identity matrix. The indices of
$\Delta W_{nn'}$ denote the perturbation site (right) and the response
site (left), respectively. For systems with a finite band gap, Ge and
Lu~\cite{GE2015} proved that the spatial distribution of $|\Delta
W_{nn'}^\pm\rangle$ decays exponentially in real space, and its
magnitude decays exponentially as the distance between sites $n$ and
$n'$.

\begin{figure}[tb]
\includegraphics[width=2.5 in]{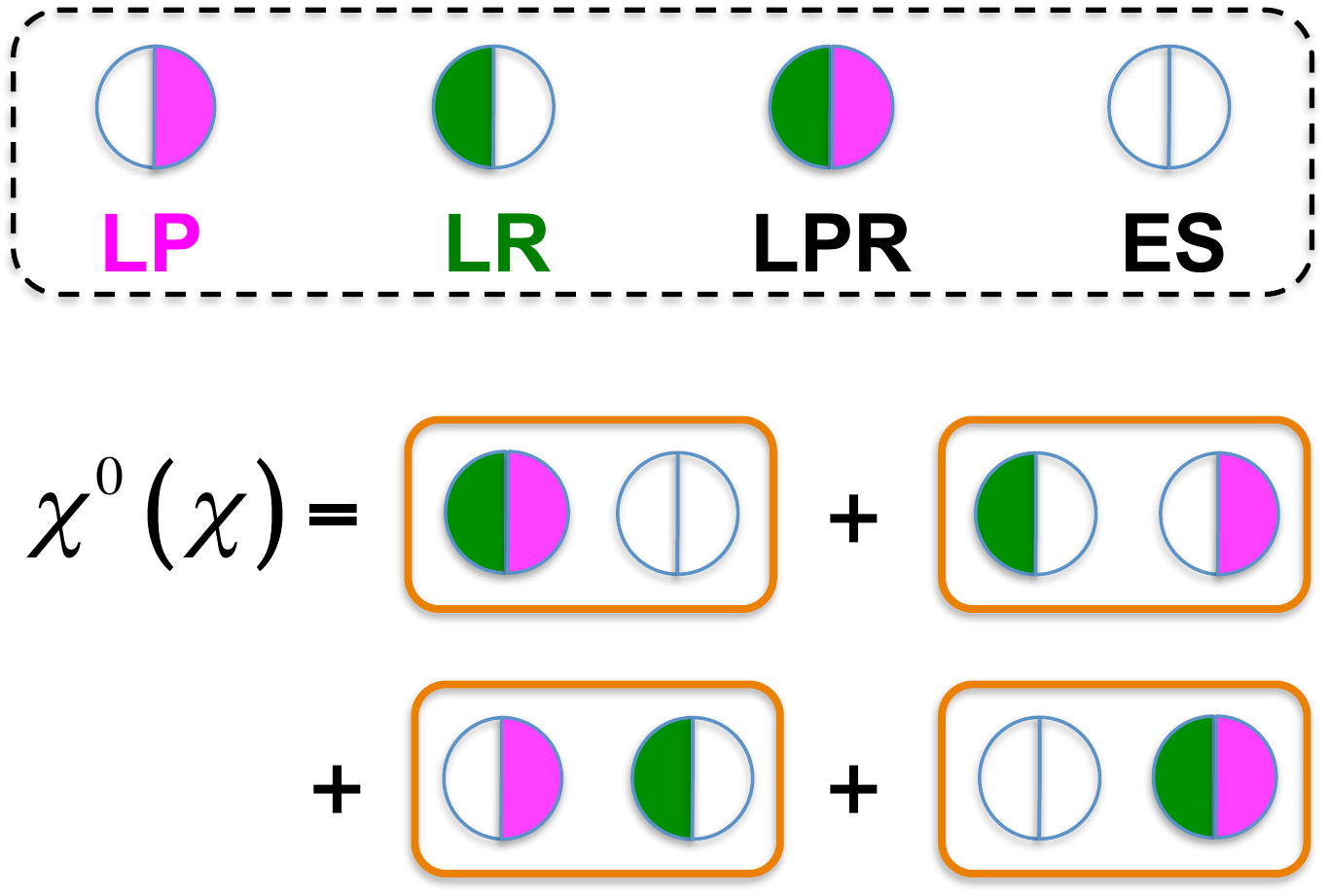}
\caption{ \label{fig-ldrt} Schematics of partitioning $\chi^0$
($\chi$) of a system with two occupied Wannier orbitals into four
partial microscopic susceptibilities. Diagrams at the top indicate all
four possible scenarios on one site: local perturbation (LP), local
response (LR), local perturbation plus local response (LPR), and an
empty site (ES).}
\end{figure}

Combining Eqs.~\ref{stern-WF} and \ref{stern-WF-sol}, one can formally
construct the partial response densities and partial microscopic
susceptibilities (PMSs) on Wannier centers according to
\begin{eqnarray}
  \Delta\rho &=& \sum_{nn'} \Delta \rho_{nn'}, \;\;\;\;\;\;\;\;\;\;\;\;\;\;\;\;\;\;\;\;\;\;\;\;\;\;\;\;\chi^0 =\sum_{nn'} \chi^0_{nn'}, \nonumber \\
  \Delta \rho_{nn'} &=& \chi^0_{nn'}\;\Delta V_{scf}= 2\, (|\Delta W_{nn'}^+ \rangle+|\Delta W_{nn'}^- \rangle) \langle W_n|. \label{Erpart}
\end{eqnarray}
Similarly, $\chi$ can also be partitioned into PMSs through the Dyson's equation, 
\begin{equation}
\chi_{nn'} = \sum_{n''}\sum_{m=0}^{\infty} \chi^0_{nn''} \left( K \, \chi^0\right)^m_{n''n'}. \label{Erpartchi}
\end{equation}
The partition of $\chi^0$ ($\chi$) into PMSs is illustrated in
Fig.~\ref{fig-ldrt} for a system with two occupied Wannier centers. On
each site, there are four possible diagrams: local perturbation (LP)
only, local response (LR) only, local perturbation plus local response
(LPR), and an empty site (ES). Therefore, $\chi^0$ ($\chi$) of the
whole system can be partitioned into four terms: two diagonal terms
(LPR on site 1 and 2) and two off-diagonal terms (LP on one site and
LR on the other site).

PMSs like $\chi_{nn'}$ are important quantities to study the
non-locality of the response functions. In particular, $\chi_{nn'}$
associates the response density at site $n$ to the external
perturbation, $\Delta V_{ext}$, at site $n'$. In the quantum chemistry
language, similar notions have been introduced based on the local
basis set formalism in the conceptual density functional theory~\cite{GEER2003},
to define the local reactivity index~\cite{TORR2009}, the atom-condensed linear response matrix~\cite{SABL2010},
and a measure of aromaticity~\cite{SABL2012,FIAS2013}. 
It is worth noting that PMSs defined through the LDRT are formally additive,
and do not suffer basis set-dependent errors in the local basis set
approach. By summing over the perturbation index, one can construct
local response density, $\Delta \rho_{n}=\sum_{n'} \Delta \rho_{nn'}$,
and local susceptibilities, $\chi^0_n=\sum_{n'}\chi^0_{nn'}$ and
$\chi_n =\sum_{n'}\chi_{nn'}$, or their symmetric form,
$\tilde{\chi}^{0}_n=\frac{1}{2}\sum_{n'}(\chi^0_{nn'}+\chi^0_{n'n})$
and $\tilde{\chi}_n=\frac{1}{2}\sum_{n'}(\chi_{nn'}+\chi_{n'n})$. A
Dyson-like relation exists between $\chi_n$ and $\chi_n^0$,
\begin{equation}
\chi_n=\chi^0_n\,(1-K\chi^0)^{-1}. \label{Echin}
\end{equation}
$\chi_n$ provides a local measure of excited
state properties projected onto a Wannier orbital,
such as the bond polarizability~\cite{GE2015}.

\subsection{Electronic density response of the local perturbation}
We start by dividing the Wannier centers of the whole system into two
sub-systems: the molecule of interest ($M$) and the environment ($E$)
as shown in Fig.~\ref{fig-lext}b. It is convenient to define local quantities
associated with each sub-system, $S$ ($S$ = $M$ or $E$): $\Delta
\rho_{S}=\sum_{n\in S} \Delta \rho_{n}$, $\chi^0_S =\sum_{n\in
S}\chi^0_{n}$, and $\chi_S =\sum_{n\in S}\chi_{n}$. A quantum
mechanical description of the molecular polarizability embedded in a
medium relies on the density response to the perturbation in the local
field, or in short, the local perturbation. Conceptually, within the
linear response theory it implies a) to freeze both the
external perturbation potential and the induced potential due to the
polarization of the environment, and b) to obtain the self-consistent
solution of the response density of the molecule.
In the same spirit, Buin and Iftimie~\cite{BUIN2009} proposed the frozen orbitals polarizability 
model within the IDM based on the MLWFs, as an alternative to Heaton \emph{et al.}'s 
formulation~\cite{HEAT2006}. In practice, one
has to ``unscreen" (US) the PMSs on the molecule ($\chi_S$) to remove the
screening effects from the environment. For this purpose, we define
the unscreened molecular susceptibility ($\chi_M^{us}$) as
\begin{equation}
\Delta \rho_{M}=\chi_M^{us}\,\Delta V_{loc},
\end{equation}
where $\Delta V_{loc}=\Delta V_{ext}+K\,\Delta \rho_{M}$ is the perturbation in the local potential.
Clearly, this quantity is different from $\chi$ that is related to
external perturbation, $\Delta V_{ext}$, and $\chi^0$ that is related to the
self-consistent field perturbation, $\Delta V_{scf}$. The unscreening procedure implies
that
\begin{equation}
\chi_M^{us}=\chi^0_M+\chi^0_M\,K\,\chi^0_M+\chi^0_M\,(K\,\chi^0_M)^2+\cdots=\chi_M^0+\chi_M^0\,K\,\chi_M^{us}. \label{Echim}
\end{equation}
The molecular polarizability is given by
\begin{equation}
\alpha_{M,ij}=-\int dr_i dr'_j\; r_i \,\chi_M^{us}
(\omega;\bm{r},\bm{r}')\,r'_j\equiv -Tr \left [ r_i (\chi_M^{us} r'_j) \right ],
\end{equation}
where $i$ and $j$ denote Cartesian axes. Because $\chi_M^0$ and
$\chi_M$ are additive, so is the effective molecular polarizability, $\alpha_{M,ij}^{eff}=-Tr \left [ r_i (\chi_M
r'_j) \right ]$.  On the other hand, because $\chi_M^{us}$ is not additive, neither is $\alpha_{M}$.

In order to reveal the relation between the above quantum mechanical
description and the IDM of Eq.~\ref{Edipole}, we
similarly define $\chi_E^{us}$ as the unscreened susceptibility of the
sub-system $E$ under the perturbation of the local field,
\begin{equation}
\Delta \rho_{E}=\chi_E^{us}\,\Delta V_{loc}.
\end{equation}
It allows us to express the unscreened and screened susceptibilities of each sub-system,
\begin{eqnarray}
\chi_S^{us}&=&\chi_S^0\,(1-K\chi_S^0)^{-1}, \nonumber \\
\chi_S&=&\chi_S^0\,(1-K \chi^0)^{-1}.
\end{eqnarray}
Employing the additivity relation, $\chi^0=\chi^0_M+\chi^0_E$, we have
\begin{eqnarray}
\chi_M&=&\chi_M^{us}+\chi_M^{us}\,K\chi_E, \label{Echimquant} \\
\chi_E&=&\chi_E^{us}+\chi_E^{us}\,K\chi_M.
\end{eqnarray}

Consider $\Delta V_{ext}(\bm{r})=E_j\,r_j$. Multiply $-r_i$ from left
and $E_jr'_j$ from right on both sides of Eq.~\ref{Echimquant} and
integrate.  It follows that
\begin{equation}
\Delta\mu_{M,i}=\alpha_{M,ij}\,E_j-Tr\left [ r_i ( \chi_M^{us}\,K\chi_E\,r'_j)\right ] E_j. \label{Emum}
\end{equation}
Ignore $K_{xc}$ in $K$, and take the dipole approximation to expand
$v_H=\sum_{N\in E} |(\bm{r}_M+\bm{r}'')-(\bm{r}_N+\bm{r}''')|^{-1}$ to
the second order of $r_{MN}$.  The only non-vanishing terms arise from
$\bm{r}''\,T_{MN}\,\bm{r}'''$ due to the charge neutrality condition,
$\int d\bm{r} \Delta \rho(\bm{r})=0$. Consequently, the second term on
the right hand side of Eq.~\ref{Emum} becomes $\sum_{N\in E}\sum_{kl}
\langle r_i| \chi_M^{us}|r''_k\rangle \,T_{MN,kl}\,\langle r'''_l
|\chi_E |r'_j \rangle \, E_j$, where $T_{MN,kl}=\partial_k\partial_l
r_{MN}^{-1}$. It is straightforward to show that
\begin{equation}
\Delta\mu_{M,i}=\alpha_{M,ij}\,E_j+\sum_{N\in E}\sum_{kl} \alpha_{M,ik} \,T_{MN,kl} \,\Delta\mu_{N,l}, \label{Edp}
\end{equation}
which reproduces Eq.~\ref{Edipole}.  Alternatively, by dividing $E_j$
on both sides of Eq.~\ref{Edp}, one obtains the IDM without explicit dependence on the external field,
\begin{equation}
\alpha_{M}^{eff}=\alpha_{M}+\sum_{N\in E}\alpha_{M} \,T_{MN} \,\alpha_{N}^{eff}.
\end{equation}
We have proved that the IDM is the classical limit of Eq.~\ref{Echimquant} under the dipole approximation.

One may also arrange the screened and unscreened PMSs in the matrix form through
\begin{equation}
\begin{pmatrix}
\chi_M \\ \chi_E
\end{pmatrix}
=\begin{pmatrix}
1 & -\chi_M^{us}\,K\\
-\chi_M^{us}\,K & 1
\end{pmatrix}^{-1}
\begin{pmatrix}
\chi_M^{us} \\ \chi_E^{us}
\end{pmatrix}. \label{Echimchie}
\end{equation}
By expanding the matrix inversion in Taylor series, the first and
second order corrections to $\chi_M$ are $\chi_M^{us}\,K\,\chi_E^{us}$
and $\chi_M^{us}\,K\,\chi_E^{us}\,K\,\chi_M^{us}$, respectively. In
the limit that $M$ and $E$ are spatially fully separated, Dobson~\cite{DOBS94}
derived the same low order correction terms to
$\chi_M$, which have been used by one of us to derive three-body terms in the RPA
correlation energy~\cite{LU2010}. Because Eq.~\ref{Echimchie} 
includes contributions at infinite orders, it is also valid in the regime where the electron
densities of $M$ and $E$ overlap. We also note the similarity between Eq.~\ref{Echimquant} in this work and Eq.~41 in the linear response theory of
subsystem TDDFT~\cite{PAVA2013,KRIS2015}.

\section{Computational Details}
In this study, ground state and linear response calculations were performed using the Perdew-Burke-Ernzerhof (PBE)~\cite{PERD96,*PERD97} 
exchange-correlation functional with the SG15 optimized norm-conserving Vanderbilt (ONCV) 
pseudopotentials~\cite{HAMA2013,SCHL2015, oncv} together with the $\Gamma$-point sampling. 
The kinetic energy cutoff of the planewave basis set  was chosen at $65$ Rydberg. All the gas phase calculations
were performed using a simple cubic supercell of $20$~\hbox{\AA}. MLWFs were constructed with Wannier90~\cite{MOST2014}.
$\alpha_M$ is calculated from the self-consistent solution of $\chi_M^{us}$ according to Eq.~\ref{Echim} 
through the DFPT implemented in a customized version of Quantum ESPRESSO~\cite{Giannozzi2009}.
At each self-consistent step, the generalized Sternheimer equations are solved, and the density response is projected
onto $M$ as $\Delta \rho_M=\chi^0_M \Delta V_{loc}$. The converged density response is used to compute $\alpha_M$.

In practice, we first compute the ground state of the system, which can be either a finite system in a supercell or an extended 
system described under the periodic boundary condition. Since we are using a $\Gamma$-point formalism for extended systems, the unit cell size has to be sufficiently large.
Next we construct Wannier orbitals for occupied states, and associate Wannier orbitals to their corresponding water molecules.
Different environmental effects can be quantified separately based on the choices of R$_1$, R$_2$ and R$_3$ denoted by 
$(n_{R_1}:n_{R_2}:n_{R_3})$ in the subsequent linear response calculations. Each of R$_1$, R$_2$ and R$_3$ denotes a subset of water molecules.
Specifically in the study of $\alpha_{\mr{H_2O}}$, R$_1$ is restricted to one water molecule, i.e. $n_{R_1}=1$. 
For the CF effect, R$_3$ contains the whole system, which means that $V_{scf}$ of the whole system is used in the Sternheimer equations.
We restrict the charge transfer effects inside the subsystem defined by R$_2$. If $n_{R_2}=n_{R_3}$, no truncation is applied, and 
the generalized Sternheimer equations are solved for all the occupied Wannier orbitals. If  $n_{R_2}\in n_{R_3}$, the generalized Sternheimer equations are 
truncated so that only the occupied Wannier orbitals corresponding to  R$_2$ are included.

The effect of the crystal field is investigated by comparing water clusters with different sizes with extended systems. To this end,
we consider two extended systems (ice I\emph{h} and liquid water) and three types of water clusters: water monomers ($n_{R_3}=1$), 
water clusters including the first solvation shell  ($n_{R_3}=5$), and water clusters including the first and second solvation 
shells ($n_{R_3}=17$). The structures of gas phase water clusters are fixed at the same geometry as those in ice I\emph{h} or liquid water
in order to eliminate the ambiguity due to the structural changes. To study the distance dependence of the CT effect, the region of R$_2$ is 
varied in size, with the lower limit being one water molecule (R$_1$) and the upper 
limit being the whole system (R$_3$).

The ice I\emph{h} structure is modeled by an orthorhombic supercell ($n_{R_3}=96$; $a=13.30$~\AA, $b=15.36$~\AA, and $c=14.47$~\AA) constructed from 
the hexagonal unit cell ($a=b=7.78$~\AA, $c=7.33$~\AA, and $\gamma=60$\hbox{\degree}) of Ref.~\onlinecite{SANT2013}. Although the effect
of proton disorder can in principle be studied by using different proton-ordered, energetically quasi-degenerate ice I\emph{h} 
structures, it is beyond the scope of the current work.
The structures of liquid water were taken from the trajectories  of the \emph{ab initio} molecular dynamics simulation of $64$-molecule water samples
at 400 K, the water PBE400 dataset~\cite{pbewater400} ($n_{R_3}=64$). The supercell size is $12.41$~\AA.
$\alpha_{\mr{H_2O}}$ are averaged over the first $20$ snapshots of the 
PBE$400$\textunderscore $64$ subset, resulting in $1280$ water molecule geometries.

\section{Results and Discussions}
\subsection{Water monomer in the gas phase}
$\alpha_{\mr{H_2O}}$ of the gas phase water monomer optimized with the PBE exchange-correlation functional is listed in Table~\ref{Talpha}.
$\bar{\alpha}_{\mr{H_2O}}=1.59$~\AA$^3$, and the three principal components are $\alpha_{xx}=1.58$~\AA$^3$, $\alpha_{yy}=1.60$~\AA$^3$, 
and $\alpha_{zz}=1.58$~\AA$^3$, respectively. The principal axes are define as the following: $x$ along the bisector direction, $y$ along the in-plane perpendicular direction, and $z$ along 
the normal direction of the molecular plane. These values are in close agreement with the previous PBE results~\cite{WAN2013}. 
$\bar{\alpha}_{\mr{H_2O}}$ overestimates the experimental value by $8$\%, in line with the error expected from PBE. This overestimation is associated with the
underestimation of band gap and band width well-known for semi-local approximations to the exchange and correlation potential and in line with Penn's model~\cite{PENN62} of
the dielectric constant. Nevertheless, our main focus is to understand the general trend of the environmental effects on $\alpha_{\rm{H_2O}}$ in the condensed phase, i.e., changes in $\alpha_{\rm{H_2O}}$, not the absolute value.

We also checked our gas phase results using the BLYP functional~\cite{BECK88, LEE88} in order to compare with the BLYP results in the literature~\cite{SALA2008, BUIN2009}.
As shown in Table~\ref{Talpha}, BLYP results in Refs.~\onlinecite{SALA2008, BUIN2009} are systematically smaller than our results,  with $\bar{\alpha}_{\mr{H_2O}}$ is about $8$\% smaller than our result. This difference is caused by the supercell size convergence issue in the finite field calculations~\cite{UMAR2003} in previous studies~\cite{SALA2008, BUIN2009}. As pointed out by Buin and Iftimie~\cite{BUIN2009}, $\alpha_{xx}$ obtained from a $10$~\hbox{\AA} cubic box is known to underestimate the fully converged value by $8$\%, and similar underestimations are likely to occur also in their liquid water calculations. Their observation is consistent with the fact that $\alpha_{xx}$ in Refs.~\onlinecite{SALA2008, BUIN2009} is about $8$\% smaller than our result. On the other hand, our linear response method does not suffer this convergence issue. For example, gas phase $\alpha_{ii}$ calculated with 20~\hbox{\AA} and 30~\hbox{\AA} cubic supercells are within 0.04\%. If we compare PBE and BLYP results in this work, $\bar{\alpha}_{\mr{H_2O}}^{\mr{BLYP}}$ is slightly larger than $\bar{\alpha}_{\mr{H_2O}}^{\mr{PBE}}$ by $1$\%.

\begin{table}[tbh]
  \caption{\label{Talpha} Molecular polarizability of water (in \AA$^3$) of the gas phase monomer, ice I\emph{h} and liquid water. 
  Results in liquid water are the statistical average ($\langle\alpha\rangle$) and the standard deviation ($\sigma$). 
  The $x$ axis is the dipole axis, and the $z$ axis is the normal direction of the molecular plane. Gas phase molecules in the rows of ice I\emph{h} and
  liquid water have the same geometry as in the condensed phase.
  CF and CT denote the crystal field and charge transfer, respectively.}
  \begin{tabular}{cc|cccccccc}
    \hline\hline
    & & $\alpha_{xx}$ & $\sigma_{xx}$ & $\alpha_{yy}$ & $\sigma_{yy}$ & $\alpha_{zz}$ &$\sigma_{zz}$ &$\bar{\alpha}_{\mr{H_2O}}$& $\sigma$\\
    \hline
   monomer  &PBE$^a$ & 1.58 &  & 1.60 &  & 1.58 & &1.59& \\
   &BLYP$^{a}$ &1.60 && 1.63 && 1.59 && 1.61 &\\
   &BLYP$^{b,c}$ & 1.47 &  & 1.53&  & 1.42 & && \\
   &exp$^d$ & $1.468\pm 0.003$ &  & $1.528\pm 0.013$ &  & $1.415\pm 0.013$ & &$1.470\pm 0.003$& \\
    \hline
   &gas phase & 1.65 &  & 1.71 &  & 1.61 & &1.66& \\  
    ice I\emph{h}&CF & 1.58& & 1.44 & & 1.76 & &1.60& \\  
    & CF+CT & 1.69 & & 1.54 & & 1.88 & & 1.70& \\
    \hline
       &gas phase & 1.64 & 0.05 & 1.69 & 0.08 & 1.61 & 0.03 & 1.65 & 0.05 \\  
    water &CF & 1.66 & 0.10 & 1.55 & 0.09 & 1.78 & 0.14 & 1.66 & 0.08 \\  
    & CF+CT & 1.76 & 0.11 & 1.63 & 0.10 & 1.86 & 0.16 & 1.75 & 0.09 \\
    & BLYP$^b$ & 1.45 &  & 1.42 &  & 1.48 &  &  &  \\
    & BLYP$^c$ & 1.44 & 0.03 & 1.41 & 0.03 & 1.49 & 0.03 & &  \\
    \hline\hline
\end{tabular}  
      \small
      \item $^a$ this work. $^b$ Ref.~\onlinecite{SALA2008}. $^c$ Ref.~\onlinecite{BUIN2009}. $^d$ Ref.~\onlinecite{MURP77}.
\end{table}

\subsection{Isotropic molecular polarizability of water}
\begin{figure}[bth]
\includegraphics[width=3.0 in]{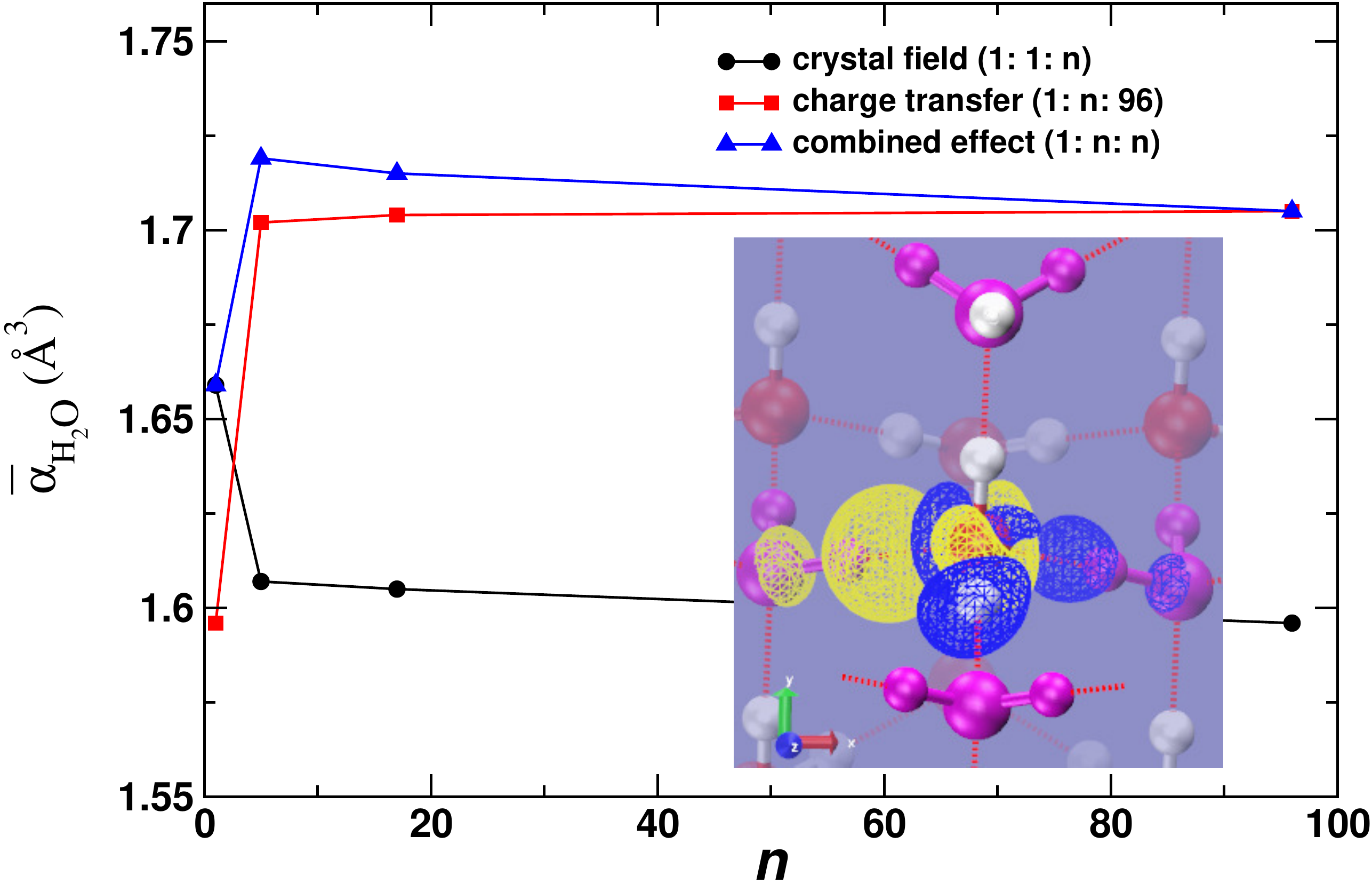}
\caption{\label{fig-ice} Environmental effects on the isotropic molecular
polarizability of water ($\bar{\alpha}_{\mr{H_2O}}$) in ice I\emph{h}. Notation 
$(n_{R_1}:n_{R_2}:n_{R_3})$ corresponds to the number of water molecules in the
molecular region ($n_{R_1}$), the charge transfer region ($n_{R_2}$) and the crystal
field region ($n_{R_3}$) as defined in Fig.~\ref{fig-lext} and Section III. Inset
shows the isosurface of the electron density response due to the charge transfer under
a uniform electric field along the $x$ axis (the red arrow), i.e., 
$\Delta\Delta\rho^{\rm{CT}}=\Delta\rho(1,96,96)-\Delta\rho(1,1,96)$. Yellow indicates
electron accumulation; blue indicates electron depletion. Water molecules are visualized in the ball-and-stick representation with 
oxygen and hydrogen atoms represented by red and white spheres. The four water molecules tetrahedrally H-bonded to the central water
molecule are highlighted in purple. The isosurface plot was generated with VMD~\cite{vmd}.}
\end{figure}

Environmental effects on $\bar{\alpha}_{\mr{H_2O}}$ in ice I\emph{h} are shown in Fig.~\ref{fig-ice}. 
 Because the geometries of each water molecule in the ground state ice I\emph{h} model are almost identical,
the computed $\bar{\alpha}_{\mr{H_2O}}$ differ by less than $0.15\%$. Therefore we present the results from a single water molecule in ice I\emph{h}.
We consider four systems: three gas phase water clusters with $n_{R_3}=1$ (monomer), $5$ (first shell), and $17$ (second shell) and ice I\emph{h} with $n_{R_3}=96$.
Here gas phase clusters refer to water molecules with the same geometry as in ice I\emph{h} to enable a straight comparison.
First, we focus on the CF effect by fixing $n_{R_2}=1$ and varying $n=n_{R_3}$. The effect of the CF is to reduce 
$\bar{\alpha}_{\mr{H_2O}}$. Compared to $\bar{\alpha}_{\mr{H_2O}}=1.66$~\AA$^3$ of the gas phase monomer, $\bar{\alpha}_{\mr{H_2O}}$ decreases by $3.1$\% caused 
by the CF of the first solvation shell. When the full CF of ice I\emph{h} is included, $\bar{\alpha}_{\mr{H_2O}}$ decreases by $3.8$\%. 
Gubskaya and Kusalik~\cite{GUBS2001} reported an opposite trend, i.e., an increase of $\bar{\alpha}_{\mr{H_2O}}$ by 
$1.2 \sim 5.3$\%, which is likely due to the approximate nature of their local field models represented by the electrostatic field of a few point charges.

Next, we analyze the CT effect in ice I\emph{h} by fixing $n_{R_3}=96$ and varying $n=n_{R_2}$. In general, we expect the CT effect to enhance $\alpha_M$,
as the excitation of neighboring water molecules has a constructive contribution.
Indeed after we include the CT effect from the first solvation shell, $\bar{\alpha}_{\mr{H_2O}}$ increases from $1.60$ to $1.70$~\AA$^3$, which is more 
significant than the total CF effect. Since the CT effect is short-ranged, it saturates readily at $n_{R_2}=5$. At $n_{R_2}=96$, $\bar{\alpha}_{\mr{H_2O}}$
further increases by only $0.003$~\AA$^3$. Finally, we combine the CF and CT effects by simultaneously varying $n=n_{R_2}=n_{R_3}$. As shown in Fig.~\ref{fig-ice},
the CT effect dominates within the first solvation shell, and $\bar{\alpha}_{\mr{H_2O}}$ reaches the maximum at $1.72$~\AA$^3$. Beyond that, the CT effect fades away, and the
tail of the CF effect takes over. $\bar{\alpha}_{\mr{H_2O}}$ decreases slowly by $0.01$~\AA$^3$ at $n=96$. Overall, $\bar{\alpha}_{\mr{H_2O}}$ in ice I\emph{h} 
is larger than the gas phase value by $2.8$\%.

The inset of Fig.~\ref{fig-ice} shows the isosurface of the electron density response in ice I\emph{h} due to the charge transfer under
a uniform electric field along the $x$ axis (the red arrow), i.e., $\Delta\Delta\rho^{\rm{CT}}=\Delta\rho(1,96,96)-\Delta\rho(1,1,96)$.
Clearly one can see a positive induced dipole moment emerging from the four nearest neighbor water molecules (highlighted in purple) that are 
H-bonded to the central water molecule. Previously, Lu \emph{et al.}~\cite{LU2008} applied the MLWF procedure to localize the 
eigenvectors of the static dielectric matrices in ice I\emph{h} and liquid water, and identified dominant screening modes that are either localized on individual water molecules 
or involving H-bonded water dimers. Based on our study, the physical origin of these modes becomes clear, which are intramolecular excitations and
intermolecular charge transfer excitations.

Next we compute $\bar{\alpha}_{\mr{H_2O}}$ in liquid water, where the thermal disorder effects play a key role in contrast to the results of ice I\emph{h} at zero temperature.
Here we divided the thermal disorder effects into intramolecular contributions on individual water molecules and intermolecular contributions from the crystal field including
H-bonds and long-range electrostatic effects. To isolate the intramolecular component, we sampled $\bar{\alpha}_{\mr{H_2O}}$ of gas phase water monomers 
with the same geometries as those in liquid water. The intermolecular contributions were extracted by comparing $\bar{\alpha}_{\mr{H_2O}}$ of the gas phase
and liquid water that includes environmental effects.

\begin{figure}[bth]
\includegraphics[width=3.0 in]{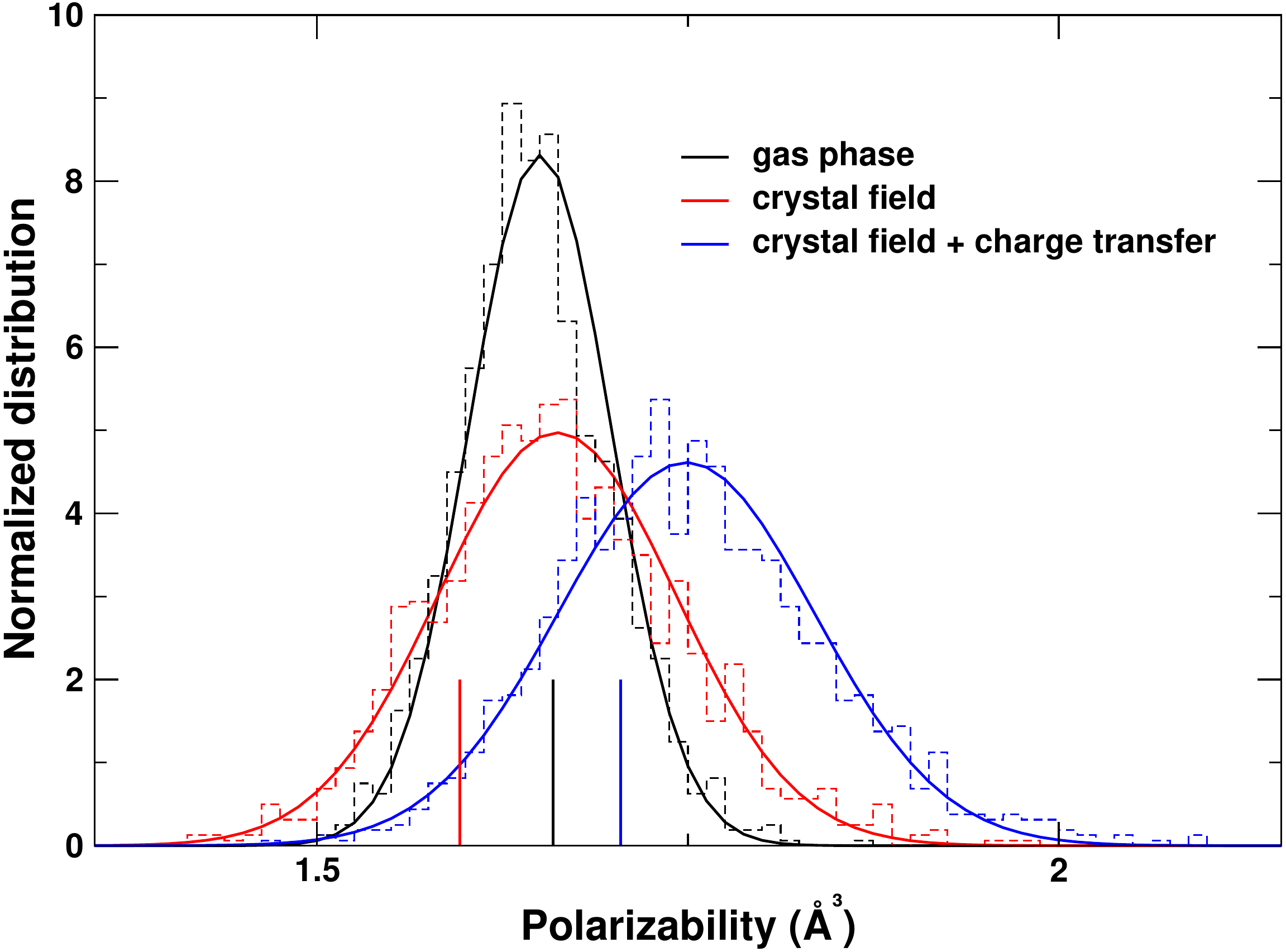}
\caption{\label{fig-water-avg} Distributions of the isotropic molecular
polarizability of water ($\bar{\alpha}_{\mr{H_2O}}$) in liquid water. Short vertical lines indicate
corresponding values in ice I\emph{h}.}
\end{figure}

As shown in Table~\ref{Talpha} and Fig.~\ref{fig-water-avg}, $\langle\bar{\alpha}_{\mr{H_2O}}\rangle$ is $1.65$~\AA$^3$ in the gas phase, which is averaged over 
$1280$ monomer configurations. Here $\langle\cdots \rangle$ denotes the statistical average. This mean value is nearly the same as that in ice I\emph{h} ($1.66$~\AA$^3$),
and the intramolecular thermal disorder in the PBE water leads to a standard deviation of $\sigma=0.05$~\AA$^3$. However, once the CF effect is included, 
$\langle\bar{\alpha}_{\mr{H_2O}}\rangle$ slightly increases to $1.66$~\AA$^3$, which is substantially larger than that in ice I\emph{h} ($1.60$~\AA$^3$) by $4$\%.
At the same time, $\sigma$ increases to $0.08$~\AA$^3$, indicating a notable intermolecular contribution. The effect of the CT in liquid water is 
similar to that in ice I\emph{h}, which further increases $\bar{\alpha}_{\mr{H_2O}}$ by $0.09$~\AA$^3$ with $\sigma$ almost unaffected. This suggests that the crystal field
is the primary source of the intermolecular thermal disorder effects.

\subsection{Anisotropic effects of the molecular polarizability of water}
The crystal field in the condensed phases can modify the the electronic properties of individual water molecules significantly.
For example, solvated water molecules form hydrogen-bonds (H-bonds) with their neighbors in a tetrahedral geometry, which has a 
strong influence on the Wannier function centers (WFCs: two from the O-H covalent bonds and two from the oxygen lone pairs).
As a consequence, the O-H bond WFCs are pulled in and the lone pair WFCs
are pulled out as compared to the gas phase monomer~\cite{SILV99}. In order to gain insights into the chemical 
origin of the CF and CT effects, we examine the anisotropic effects in ice I\emph{h} and liquid water by
decomposing the $\alpha_{\mr{H_2O}}$ tensor onto three internal principal axes, and correlating the changes in the molecular polarizability 
($\Delta\alpha_{ii},\, i=x,y,z$) with the changes in the O-WFC pair correlation function, $g(r)$, in terms of $\Delta r_{\mr{O-WFC}}$.

\begin{figure}[tbh]
\includegraphics[width=3.0 in]{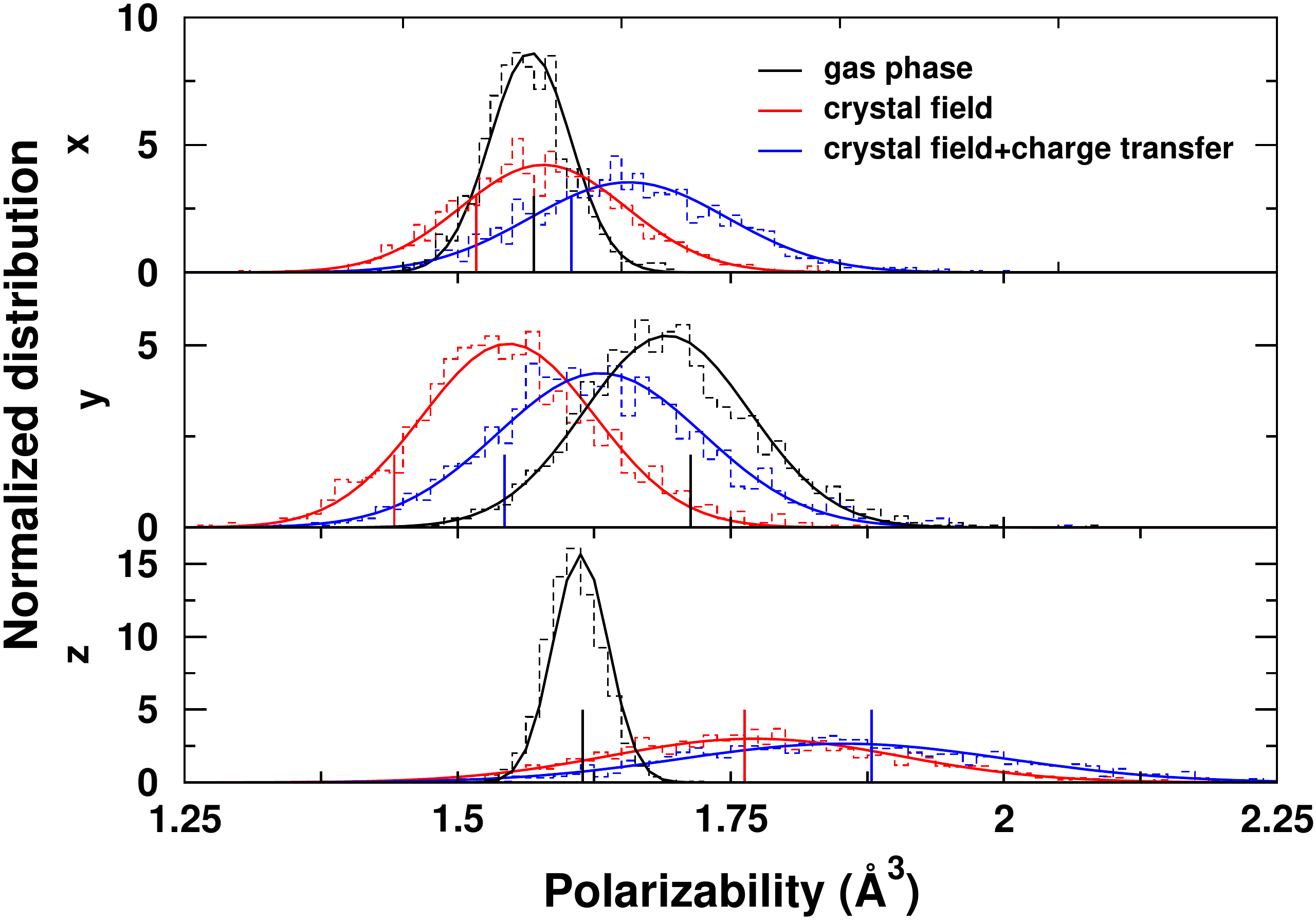}
\caption{\label{fig-water} Distributions of the molecular
polarizability of water in liquid water along three principal
directions ($x$: bisector direction, $y$: in-plane perpendicular
direction, and $z$: out-of-plane direction). Short vertical lines indicate
corresponding values in ice I\emph{h}.}
\end{figure}

As shown in Table~\ref{Talpha}, the CF of ice I\emph{h} causes $\alpha_{xx}$
and $\alpha_{yy}$ to decrease by $4$\% and $16$\%, and $\alpha_{zz}$ to increase by $9$\%. It appears that the in-plane suppression 
effect is more pronounced in the $y$ direction. Interestingly, this anisotropic CF effect can be qualitatively captured using the point 
charge model~\cite{GUBS2001} except for the $x$ component, which yields $\Delta\alpha_{xx}=4\%$, $\Delta\alpha_{yy}=-4\%$, 
and $\Delta\alpha_{zz}=16\%$ from their model II. This strong anisotropic CF effect is thus characteristic of the H-bond network in the condensed phase. 
On the other hand, the effect of the CT is mostly isotropic, increasing each component by about $0.1$~\AA$^3$.\

Similar to the isotropic molecular polarizability, $\langle \alpha_{ii}\rangle$ of gas phase monomers
in liquid water ($1.64$, $1.69$ and $1.61$~\AA$^3$) are very close to those in ice I\emph{h} ($1.65$, $1.71$ and $1.61$~\AA$^3$) 
as shown in Table~\ref{Talpha} and Fig.~\ref{fig-water},
with the largest deviation of $1$\% from the $y$ direction. The largest spread is also found in the $y$ direction ($\sigma_{yy}=0.08$~\AA$^3$), 
followed by $x$ ($\sigma_{xx}=0.05$~\AA$^3$) and $z$ ($\sigma_{zz}=0.03$~\AA$^3$). The CF in liquid water also exhibits a significant anisotropic effect, which 
leads to $\Delta\langle \alpha_{ii}\rangle= +1$\%, $-9$\%, and $+11$\% as compared to the gas phase values. Another important feature is that the $z$ component acquires the largest
spread ($\sigma_{zz}=0.14$~\AA$^3$), and the spreads in $x$ and $y$ directions are much smaller ($\sigma_{xx}=0.10$~\AA$^3$ and $\sigma_{xx}=0.09$~\AA$^3$).
The CT effect in liquid water increases $\langle \alpha_{ii}\rangle$ almost uniformly by $0.1$~\AA$^3$, and the spreads are only slightly increased.

\begin{figure}[bth]
\includegraphics[width=3.0 in]{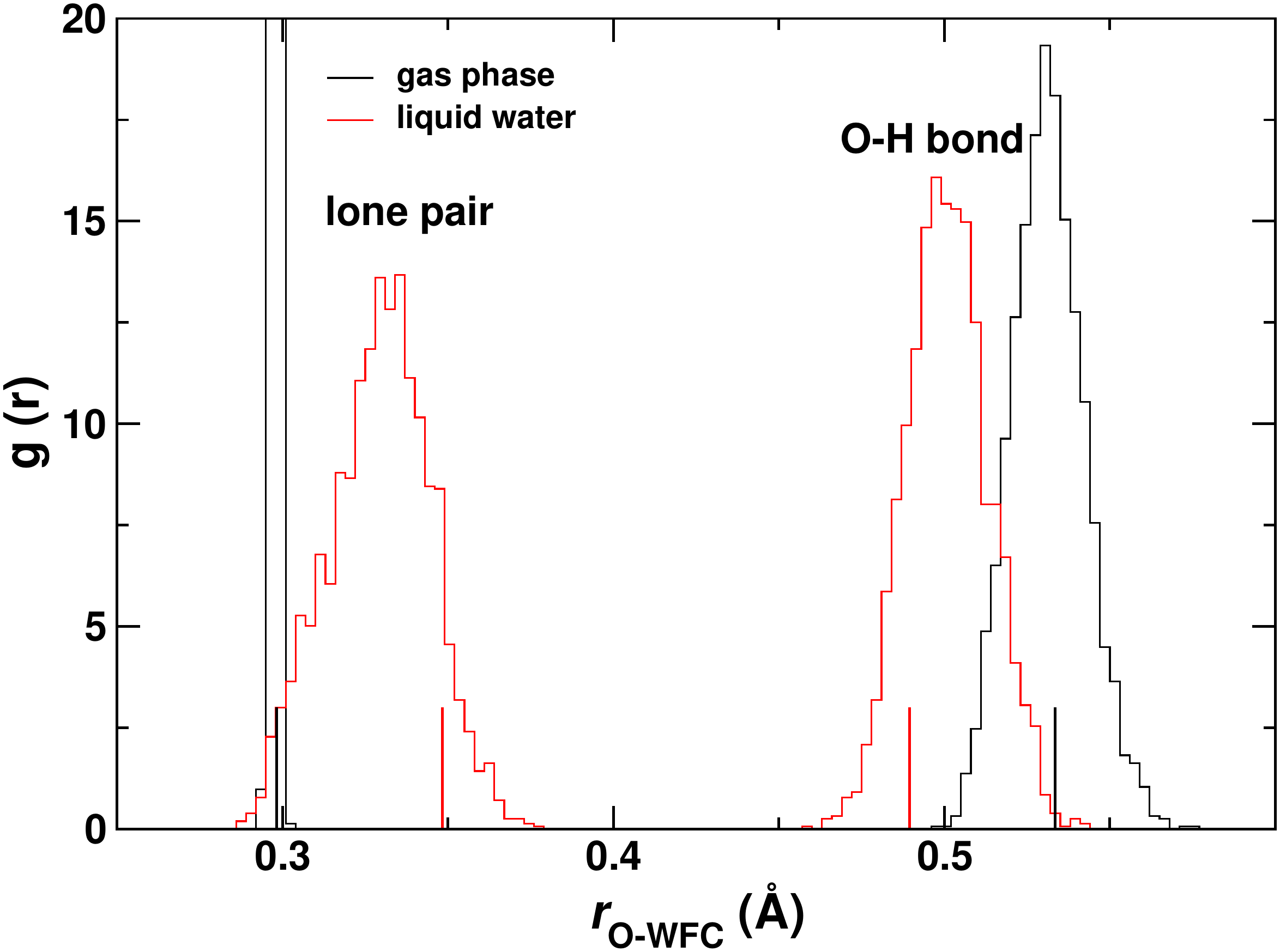}
\caption{\label{fig-wannier} Pair correlation function of the oxygen - Wannier function centers (O-WFCs) in liquid water. 
Short vertical lines indicate corresponding values in ice I\emph{h}.}
\end{figure}

The origin of the CF effects in ice I\emph{h} and liquid water was investigated using $\Delta r_{\mr{O-WFC}}$, which are the differences of O-WFC distances between the condensed phases
and the gas phase. As shown in Fig.~\ref{fig-wannier}, in ice I\emph{h} $\Delta r_{\mr{O-WFC}}=0.05$ and $-0.04$~\hbox{\AA} for the oxygen lone pair and the O-H bond, respectively.
In other words, O-H bond WFCs are pulled in from $0.53$~\hbox{\AA} by $0.04$~\AA, while lone pair Wannier
centers are pulled out from $0.30$~\hbox{\AA} by $0.05$~\hbox{\AA} as compared to the gas phase monomer, in line with the established
trend in liquid water~\cite{SILV99}. The formation of the H-bonds therefore
weakens the intramolecular O-H bond, making it more ionic, and loosens the oxygen lone pairs.
Since the O-H WFs become more tightly bound to the oxygen atom, the in-plane components of $\alpha_{\mr{H_2O}}$ are suppressed.
In contrast, since the lone pair WFs become more loosely bound to the oxygen atom,  
the out-of-plane component is enhanced. In liquid water, both changes get smaller, i.e. $\Delta\langle r_{\mr{O-WFC}}\rangle=0.03$ and $-0.03$~\hbox{\AA}, 
indicating overall softer H-bonds than ice I\emph{h}. Consequently, the CF effects have a smaller impact on the in-plane components 
of liquid water ($\Delta\langle\alpha_{xx}\rangle=1$\% and $\Delta\langle\alpha_{yy}\rangle=-9$\%) than ice I\emph{h} ($\Delta\alpha_{xx}=-4$\% and $\Delta\alpha_{yy}=-16$\%),
while $\Delta \langle\alpha_{zz}\rangle$ in both systems are comparable (11\% and 9\%). The oxygen lone pair has a larger spread ($\sigma_{\rm{O-WFC}}=0.015$~\hbox{\AA}) than the O-H bond ($\sigma_{\rm{O-WFC}}=0.012$~\hbox{\AA}), which is consistent with the largest spread in $\alpha_{zz}$ (see Fig.~\ref{fig-water} and Table~\ref{Talpha}). 

\begin{figure}[bth]
\includegraphics[width=3.0 in]{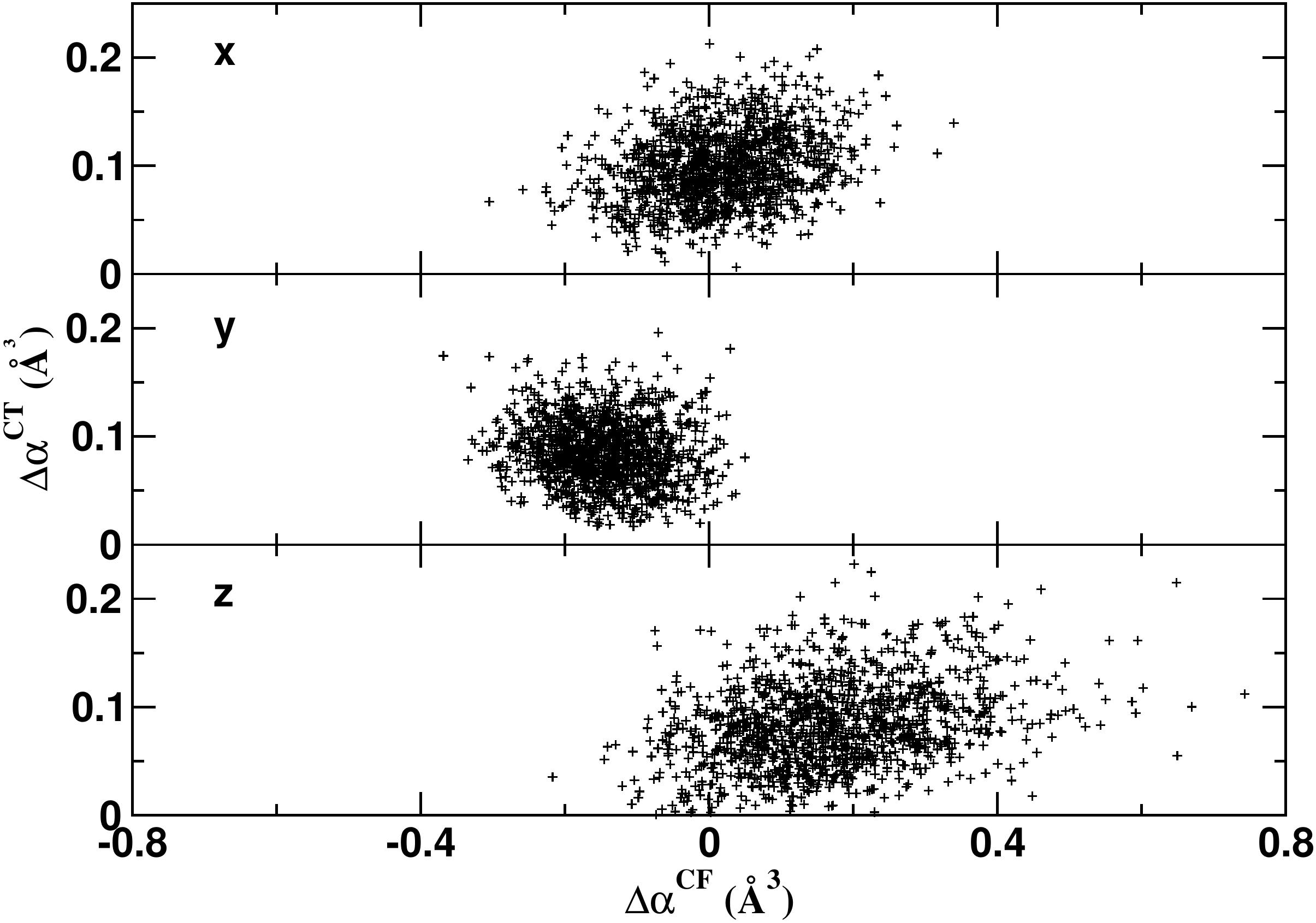}
\caption{\label{fig-delta} Correlation between the crystal field (CF)
and charge transfer (CT) contributions in the molecular polarizability
of water in liquid water.}
\end{figure}

To investigate any possible correlation between the CF and CT effects in liquid water, we track $\Delta\alpha_{ii}$ from CF and CT effects for each water molecule. As shown in
Fig.~\ref{fig-delta}, while the distribution of $\Delta\alpha_{xx}^{\mr{CF}}$ is nearly symmetric around zero, the signs of $\Delta\alpha_{yy}^{\mr{CF}}$ and $\Delta\alpha_{zz}^{\mr{CF}}$
are opposite, with one being predominantly negative and the other positive. No apparent correlation was found for the in-plane components, while there is a weak correlation between $\Delta\alpha_{zz}^{\mr{CF}}$ and $\Delta\alpha_{zz}^{\mr{CT}}$. Therefore our procedure of treating CF and CT effects separately is justified.

The main differences between our model and previous models can be understood as the following. In the point charge model~\cite{GUBS2001}, the CF is over-simplified. In the IDM, although the full CF is included at the ground state level, the Coulomb kernel is approximated by the dipole interaction.
It is not straightforward to compare the absolute values of $\alpha_{\rm{H_2O}}$ of this work with the IDM results in the literature~\cite{SALA2008,BUIN2009}, because different exchange-correlation functionals (PBE and BLYP) were used, and results in previous studies are likely not fully converged. A likely meaningful comparison is the trend of the environmental effects on $\alpha_{\rm{H_2O}}$ with respect to the gas phase values computed from the optimized geometry. In this work, the relative changes of $\langle\alpha_{ii}\rangle$ are 11\%, 6\% and 18\% in $x$, $y$, and $z$, respectively. The trends are qualitatively different from the IDM that yields -2$\sim$-1\%, -8$\sim$-7\% and 4$\sim$5\%, respectively~\cite{SALA2008,BUIN2009}. Besides the effects from different functionals, these differences may be attributed to several limitations in the IDM, such as the lack of the finite size effect and the use of the dipole-dipole approximation.

In summary, we have developed a fully \emph{ab initio} method to compute the electron density response to the perturbation in the local field based on the local dielectric response theory.
We applied this method to compute the molecular polarizability of water in condensed phases. Using the same molecular geometries as
in the condensed phases, we found that the effects of the crystal field is to reduce $\alpha_{yy}$ and enhance $\alpha_{zz}$, and that the charge transfer effect increases all the principal components uniformly. Our study provides a rigorous theoretical framework to determine  $\alpha_{\rm{H_2O}}$, essential to both the physical understanding of water and computer simulations using polarizable force field models. As the electron-based spectroscopy techniques, e.g., electron energy loss spectroscopy (EELS), has reached subangstrom spatial resolution~\cite{KAPE2015}, we expect these experimental techniques can provide more details about the microscopic dielectric response of water in the future.

We thank Mark Hybertsen, Xifan Wu, Roberto Car, Annabella Selloni, Philip B. Allen and Michele Pavanello for their helpful discussions.
This research used resources of the Center for Functional Nanomaterials, which is a U.S. DOE Office of Science Facility, 
at Brookhaven National Laboratory under Contract No. DE-SC0012704. This research used resources of the National Energy Research Scientific Computing Center, a DOE Office of Science User Facility supported by the Office of Science of the U.S. Department of Energy under Contract No. DE-AC02-05CH11231.

\clearpage
\bibliographystyle{apsrev4-1}

\end{document}